\documentstyle[12pt,epsf]{article}

\newcommand{\beq}{\begin{equation}}
\newcommand{\eeq}{\end{equation}}
\newcommand{\bea}{\begin{eqnarray}}
\newcommand{\eea}{\end{eqnarray}}
\newcommand{\rar}{\rightarrow}

\newcommand{\lan}{\langle}
\newcommand{\ran}{\rangle}
\newcommand{\bbeta}{\mbox{\boldmath$\beta$}}

\begin{document}

\font\fortssbx=cmssbx10 scaled \magstep2
\hbox to \hsize{
\includegraphics{uwlogo.ps}
\hskip.5in \raise.1in\hbox{\fortssbx University of Wisconsin - Madison}
\hfill$\vcenter{\hbox{\bf MADPH-95-899}
            \hbox{July 1995}}$ }
\vskip 2cm
\begin{center}
\Large
{\bf Modelling form factors in HQET} \\
\vskip 0.5cm
\large
  Sini\v{s}a Veseli  and M. G. Olsson  \\
\vskip 0.1cm
{\small \em Department of Physics, University of Wisconsin, Madison,
	\rm WI 53706}
\end{center}
\thispagestyle{empty}
\vskip 0.7cm

\begin{abstract}
We present a simple and straightforward method
for relating the form factors in HQET, as defined by the
covariant trace formalism, to the overlaps of the rest frame
wave functions of the light degrees of freedom. We also point
out several inconsistencies present in recent
calculations of the radiative rare $B$ decays, and also show
how these can be fixed even within the framework of the
non-relativistic quark model.
\end{abstract}

\newpage

\section{Introduction}
It is by now a well established fact that hadronic systems
containing a single heavy quark ($m_{Q}\gg\Lambda_{QCD}$)
 admit additional symmetries
which are not present in the full QCD Lagrangian \cite{isgur}.
The light degrees of freedom (LDF) in such hadrons typically have
four-momenta small compared with the heavy quark mass.
For these systems it is appropriate to adopt
 an effective theory (HQET) in which the heavy quark mass goes to
infinity, with its four-velocity fixed \cite{georgi,reviews}.
Since an infinitely massive heavy quark does not
recoil from the emission and absorption of soft ($E\approx \Lambda_{QCD}$)
 gluons, and since magnetic interactions of such a
quark are negligible ($\sim \frac{1}{m_{Q}}$), the strong interactions
of the heavy quark are independent of its mass and spin. Because of this,
HQET leads to relations between different
form factors describing
transitions in which a hadron containing a heavy quark $Q$ and
moving with four-velocity $v^{\mu}$, decays into another hadron
containing a heavy quark $Q'$, and moving with four-velocity $v'^{\mu}$.
In this way the number of independent form factors for these
decays is significantly reduced. For example, the six form factors
describing semileptonic decays $B\rar D^{(*)}e\bar{\nu}_{e}$
are, in the heavy quark limit, reduced to a single unknown form
factor, the Isgur-Wise function (IW) $\xi(v\cdot v')$.

Since the unknown Lorentz invariant form factors describing
a particular decay cannot be calculated
   from  first principles,
one still has to rely on some model of strong interactions
in order to estimate them.
In general these form factors will be related to the overlaps of
 the wave functions of the LDF
in the hadrons before and after the decay.
However, one has to be careful
in identifying form factors directly with
the overlap of the two wave functions. Depending
on the definition of the particular form factor,
there may be a coefficient of proportionality involved.
If these coefficients are not taken into account
 significantly
incorrect results can be obtained  no
matter which model for the wave functions one uses

In \cite{zalewski} it was observed that the
heavy quark limit implies a simple formula for the
wave function of any particle containing one heavy
quark. Based on this
we offer a straightforward method for relating the form factors,
as defined within the framework of the trace formalism
\cite{georgi2,korner,falk}, to overlaps
of wave functions of  the LDF before and after the decay. Even though
we are interested  here only in mesons, it
is obvious that an analogous calculation can be easily
done for baryons.  The paper is organized as
follows: in Section \ref{covariant} we review basic definitions
of the covariant trace formalism. Section \ref{method} describes
how one can easily identify the form factors in terms of
overlaps of the LDF.
We also give results for the several cases of interest
(transitions of $0^{-}$ state into excited states).
As a simple application of the results of this paper
 we consider  the non-relativistic
quark model in Section \ref{qmodel}. Among
other things, we also make a few comments
 on some recent calculations of rare $B$
decays into $K$-resonances \cite{mannel,ahmady}.
Our conclusions are summarized in Section \ref{conc}.

\section{Covariant representation of states}
\label{covariant}

The counting of the number of independent form factors is most
conveniently done within the framework of the trace formalism,
which was formulated in \cite{georgi2,korner} and generalized
to  excited states in \cite{falk}. Following \cite{falk}, and
using notation of \cite{mannel},
the lowest lying mesonic states can be described as follows:
\bea
C(v)& =& \frac{1}{2}\sqrt{m} (\not{v}+1)\gamma_{5}\ ,
\hspace*{+4.35cm}\ J^{P}=0^{-}\ ,
\ j=\frac{1}{2}\ ,\label{st1}\\
C^{*}(v,\epsilon)& =& \frac{1}{2}\sqrt{m} (\not{v}+1)\not{\epsilon}
\ ,
\hspace*{+4.3cm}\ J^{P}=1^{-}\ ,
\ j=\frac{1}{2}\ , \\
E(v)& =& \frac{1}{2}\sqrt{m} (\not{v}+1)\ ,
\hspace*{+4.65cm}
\ J^{P}=0^{+}\ ,
\ j=\frac{1}{2}\ ,\label{st3}\\
E^{*}(v,\epsilon)& =& \frac{1}{2}\sqrt{m} (\not{v}+1)
\gamma_{5}\not{\epsilon}
\ ,
\hspace*{+3.85cm}\ J^{P}=1^{+}\ ,
\ j=\frac{1}{2}\ ,\\
F(v,\epsilon)& =& \frac{1}{2}\sqrt{m}
\sqrt{\frac{3}{2}} (\not{v}+1)\gamma_{5}[
\epsilon^{\mu}-\frac{1}{3}\not{\epsilon}(\gamma^{\mu}-v^{\mu})]
\ ,
\hspace*{+0.1cm}
\ J^{P}=1^{+}\ ,
\ j=\frac{3}{2}\ ,\label{st5}\\
F^{*}(v,\epsilon)& =& \frac{1}{2}\sqrt{m} (\not{v}+1)
\gamma_{\nu}\epsilon^{\mu\nu}
\ ,
\hspace*{+3.65cm}
\ J^{P}=2^{+}\ ,
\ j=\frac{3}{2}\ ,\\
G(v,\epsilon)& =& \frac{1}{2}\sqrt{m}
\sqrt{\frac{3}{2}} (\not{v}+1)[
\epsilon^{\mu}-\frac{1}{3}\not{\epsilon}(\gamma^{\mu}+v^{\mu})]
\ ,
\hspace*{+0.6cm}
\ J^{P}=1^{-}\ ,
\ j=\frac{3}{2}\ ,\\
G^{*}(v,\epsilon)& =& \frac{1}{2}\sqrt{m} (\not{v}+1)
\gamma_{5}\gamma_{\nu}\epsilon^{\mu\nu}
\ ,
\hspace*{+3.4cm}\ J^{P}=2^{-}\ ,
\ j=\frac{3}{2}\ .\label{st8}
\eea
Here $m$ and $v$ are the mass and the four-velocity of the heavy meson,
and $j$ denotes the total spin of the LDF.
Also, $\epsilon^{\mu}$ is the polarization vector for spin $1$ states
(satisfying $\epsilon\cdot v=0$), while the tensor $\epsilon^{\mu\nu}$
describes a spin 2 object ($\epsilon^{\mu\nu}=\epsilon^{\nu\mu}\ ,\
\epsilon^{\mu\nu}v_{\nu}=0\ ,\ \epsilon^{\mu}_{\ \mu}=0$).
These eight states form four doublets: ($C$, $C^{*})$ is the
 $L=0$ doublet, ($E$, $E^{*}$) and  ($F$, $F^{*}$) are the two
$L=1$ doublets, and ($G$, $G^{*}$) is the $L=2$ doublet.

Matrix elements of a heavy quark current $J(q)=\bar{Q'}\Gamma Q$
between the physical meson states can be calculated easily
by taking the trace ($\omega=v\cdot v'$),
\beq
\lan \Psi'(v')|J(q)|\Psi(v)\ran =  {\rm Tr}[\bar{M'}(v')\Gamma M(v)]
{\cal M}(\omega)\ ,\label{tr}
\eeq
where $M'$ and $M$ denote appropriate matrices from
(\ref{st1})-(\ref{st8}), $\bar{M}=\gamma^{0}M^{\dag}\gamma^{0}$, and
${\cal M}(\omega)$ represents the LDF.
Again following \cite{falk,mannel}, we define the IW functions
for the transitions of a $0^{-}$ ground state into an excited state by
\beq
{\cal M}(\omega)
=
\left\{
\begin{array}{ll}
 \xi_{C}(\omega)\ , &
0^{-}_{\frac{1}{2}}\rar(0^{-}_{\frac{1}{2}},
1^{-}_{\frac{1}{2}})\ , \\
 \xi_{E}(\omega)\ , &
 0^{-}_{\frac{1}{2}}\rar(0^{+}_{\frac{1}{2}},
1^{+}_{\frac{1}{2}})\ ,\\
\xi_{F}(\omega)v_{\mu}\ ,&
0^{-}_{\frac{1}{2}}\rar(1^{+}_{\frac{3}{2}},
2^{+}_{\frac{3}{2}})\ ,\\
\xi_{G}(\omega)v_{\mu}\ ,&
 0^{-}_{\frac{1}{2}}\rar(1^{-}_{\frac{3}{2}},
2^{-}_{\frac{3}{2}})\ .
\end{array}\right.
\label{xie}
\eeq
The vector index in the last two definitions will be contracted
with the one
in the representations of excited states (\ref{st5})-(\ref{st8}).
We now show how one can define the IW functions given above
in terms of the overlaps of the wave functions of the initial
and final states of the LDF.

\section{Defining IW functions}
\label{method}

It has been pointed out
\cite{zalewski} that the assumption of the heavy quark limit implies
a simple formula for the wave function
of any particle containing one very heavy quark (with total angular
momentum $J$ and its projection $\lambda$),
\beq
\Psi_{J\lambda}^{(\alpha)}(v)=\sum_{\lambda_{j},\lambda_{Q}} \lan
j,\lambda_{j};\frac{1}{2},\lambda_{Q}|J\lambda\ran
\Phi_{j\lambda_{j}}(v)u_{\lambda_{Q}}(v)\ .\label{wf}
\eeq
In this formula $(\alpha)$ refers to all other quantum numbers of
the meson,
  $u_{\lambda_{Q}}(v)$ is the free Dirac bispinor
describing a heavy quark with spin $\frac{1}{2}$,
helicity $\lambda_{Q}$, and velocity $v$ (and
normalized to $\bar{u}u=2m$).
$\Phi_{j\lambda_{j}}$ is the wave function of the LDF
 with  total angular momentum $j$ (in the rest frame of
the particle), and its projection $\lambda_{j}$. For a meson,
this is the wave function of the light antiquark.

{}From (\ref{wf}) it can be easily seen that matrix elements of
the heavy quark currents $J(q)=\bar{Q'}\Gamma Q$ are linear combinations
of matrix elements
\beq
\lan \Phi'_{j'\lambda_{j'}}(v')|\Phi_{j\lambda_{j}}(v)\ran
\bar{u'}_{\lambda_{Q'}}(v')\Gamma u_{\lambda_{Q}}(v)\ .
\eeq
For a given  $\Gamma$, $\bar{u'}\Gamma u$ is a product of known
matrices, and therefore all the unknown dynamics is contained
in the  overlaps of LDF wave functions $\lan \Phi'|\Phi\ran$.
We choose  the spin projection axis ($z$) as the
velocity direction of meson $\Psi'$ as seen in the rest
frame of meson $\Psi$. From the independence of the
overlap on the direction of the $x$-axis we then have
\beq
\lan \Phi'_{j'\lambda_{j'}}|\Phi_{j\lambda_{j}}\ran = 0\ ,
\ {\rm if\ } \lambda_{j'}\not{\hspace{-1.2mm}=}\lambda_{j}\ .
\label{xaxis}
\eeq
Similarly, from the independence of the overlap on the orientation
of the $y$ axis,
\beq
\lan \Phi'_{j'\lambda_{j'}}|\Phi_{j\lambda_{j}}\ran =
\eta\eta'(-1)^{j'-j}
\lan \Phi'_{j',-\lambda_{j'}}|\Phi_{j,-\lambda_{j}}\ran\ ,
\label{yaxis}
\eeq
where $\eta$ and $\eta'$ are the orbital parities
of the initial and final state of the LDF.
In the case of interest to us, for the change of the orbital
angular momentum from $L$ to $L'$, $\eta\eta'=(-1)^{L+L'}$. Equations
(\ref{xaxis}) and (\ref{yaxis}) have been given in \cite{zalewski}.

As an illustrative example, we choose
the $0^{-}\rar0^{+}$ transitions and axial-vector current ($\Gamma
=\gamma^{\mu}\gamma_{5}$). From (\ref{wf}) we have
\beq
\Psi_{0}^{(\pm)}= \frac{1}{\sqrt{2}}[
\Phi_{\frac{1}{2},\frac{1}{2}}^{(\pm)}u_{-\frac{1}{2}}
-\Phi_{\frac{1}{2},-\frac{1}{2}}^{(\pm)}u_{\frac{1}{2}}]\ ,
\eeq
where $+$ and $-$ refer to $0^{+}$ and $0^{-}$ states, respectively.
Also, from (\ref{xaxis}) and (\ref{yaxis}) one can see that
\beq
\lan \Phi'^{(+)}_{\frac{1}{2},-\frac{1}{2}}(v')|
\Phi^{(-)}_{\frac{1}{2},-\frac{1}{2}}(v)\ran =
- \lan \Phi'^{(+)}_{\frac{1}{2},\frac{1}{2}}(v')|
\Phi^{(-)}_{\frac{1}{2},\frac{1}{2}}(v)\ran\ ,
\eeq
and all other overlaps are zero.
Therefore, it immediately follows that
\beq
\lan 0^{+},v'|\Gamma | 0^{-},v\ran =
\frac{1}{2}[\bar{u'}_{-\frac{1}{2}}\Gamma u_{-\frac{1}{2}}
-\bar{u'}_{\frac{1}{2}}\Gamma u_{\frac{1}{2}}]
\lan \Phi'^{(+)}_{\frac{1}{2},\frac{1}{2}}(v')|
\Phi^{(-)}_{\frac{1}{2},\frac{1}{2}}(v)\ran\ .
\label{g3g5}
\eeq
Now  choosing $\Gamma = \gamma^{3}\gamma_{5}$ and evaluating
(\ref{g3g5}) in the rest frame of $0^{-}$ meson, where
$v^{\mu} = (1,0,0,0)$ and $v'^{\mu} = (\omega, 0,0,
\sqrt{\omega^{2}-1})$, one easily obtains
\beq
\lan 0^{+},v'|\gamma^{3}\gamma_{5} | 0^{-},v\ran
= -\sqrt{mm'} \sqrt{2} \sqrt{\omega+1}
\lan \Phi'^{(+)}_{\frac{1}{2},\frac{1}{2}}(v')|
\Phi^{(-)}_{\frac{1}{2},\frac{1}{2}}(v)\ran\ .\label{wfap}
\eeq
On the other hand, using (\ref{st1}), (\ref{st3}) and (\ref{xie}) inside
(\ref{tr}) one finds
\beq
\lan 0^{+},v'|\gamma^{\mu}\gamma_{5} | 0^{-},v\ran
= \sqrt{mm'} (-v^{\mu}+v'^{\mu})\xi_{E}(\omega)\ ,
\eeq
which specialized to the rest frame of $0^{-}$ state yields
\beq
\lan 0^{+},v'|\gamma^{3}\gamma_{5} | 0^{-},v\ran
= \sqrt{mm'} \sqrt{\omega^{2}-1}\xi_{E}(\omega)\ .\label{trap}
\eeq
Comparing (\ref{wfap}) and (\ref{trap}) we obtain (apart from the
irrelevant overall sign)
\beq
\xi_{E}(\omega) = \sqrt{\frac{2}{\omega-1}}
\lan \Phi'^{(+)}_{\frac{1}{2},\frac{1}{2}}(v')|
\Phi^{(-)}_{\frac{1}{2},\frac{1}{2}}(v)\ran\ .
\eeq

Of course, in order to obtain this formula for $\xi_{E}$
(with the same overall sign),
we could have chosen  any other
component and any other current giving a non-vanishing matrix element.
Also, instead of $0^{-}\rar 0^{+}$ transitions, we
could have chosen $0^{-}\rar 1^{+}$ transitions, and specialize
to any of the three possible polarizations of the $1^{+}$ state.
Finally, any other reference frame besides the rest frame
should yield the same expression.

Let us summarize the results obtained following the
simple proceedure outlined above for  several cases of interest.
We emphasize that all the results given here
were explicitly verified
for the choices of $\Gamma = \gamma^{\mu},
\gamma^{\mu}\gamma_{5},\gamma^{\mu}\gamma^{\nu},
\gamma^{\mu}\gamma^{\nu}\gamma_{5}$, and  in  two
convenient reference frames:
besides the rest frame
of the $0^{-}$ meson, we have  also used the
Breit frame $({\bf v}=-{\bf v}')$, in which
$v^{\mu} = (\sqrt{\frac{\omega+1}{2}},0,0,-\sqrt{\frac{\omega-1}{2}})$
and
$v'^{\mu} = (\sqrt{\frac{\omega+1}{2}},0,0,\sqrt{\frac{\omega-1}{2}})$.
Polarization vectors describing spin $1$ states ($\epsilon\cdot v'=0$)
 were the standard
ones,
$\epsilon^{(\pm)} = \mp \frac{1}{\sqrt{2}}(0,1,\pm i,0)$ in both frames,
$\epsilon^{(0)} = (\sqrt{\omega^{2}-1},0,0,\omega)$
in the rest frame of $0^{-}$, and
$\epsilon^{(0)} =
(\sqrt{\frac{\omega-1}{2}},0,0,\sqrt{\frac{\omega+1}{2}}) $
in the Breit frame. Let us also, for the sake of simplicity,  define
\beq
\lan \Phi'|\Phi\ran \equiv
\lan \Phi'^{(\alpha')}_{j',\frac{1}{2}}(v')|
\Phi^{(\alpha)}_{\frac{1}{2},\frac{1}{2}}(v)\ran\ ,
\eeq
and state our results in terms of this overlap:
\begin{itemize}
\item $0_{\frac{1}{2}}^{-}\rar (0_{\frac{1}{2}}^{-},
1_{\frac{1}{2}}^{-})$ transitions.
In this case we obtain
\beq
\xi_{C}(\omega) =
\sqrt{\frac{2}{\omega+1}}\lan \Phi'|\Phi\ran \ .\label{xic}
\eeq
This expression was obtained in \cite{zalewski2}.
\item$0_{\frac{1}{2}}^{-}\rar (0_{\frac{1}{2}}^{+},
1_{\frac{1}{2}}^{+})$ transitions.
Here, as shown in the previous section,
\beq
\xi_{E}(\omega) =
\sqrt{\frac{2}{\omega-1}}\lan \Phi'|\Phi\ran \ .
\eeq
\item$0_{\frac{1}{2}}^{-}\rar (1_{\frac{3}{2}}^{+},
2_{\frac{3}{2}}^{+})$ transitions.
In this case we have
\beq
\xi_{F}(\omega) =
\sqrt{\frac{3}{\omega-1}}\frac{1}{\omega+1}\lan \Phi'|\Phi\ran \ .
\label{fxi}
\eeq
\item$0_{\frac{1}{2}}^{-}\rar (1_{\frac{3}{2}}^{-},
2_{\frac{3}{2}}^{-})$ transitions.
Here,
\beq
\xi_{G}(\omega) =
\sqrt{\frac{3}{\omega+1}}\frac{1}{\omega-1}\lan \Phi'|\Phi\ran \ .
\label{xig}
\eeq
\end{itemize}

We now proceed to relate the overlaps $\lan \Phi'|\Phi\ran$
to the actual wave functions describing the LDF
in the rest frame of the particle.
 Following \cite{zalewski2,hogaasen},
in the valence quark approximation,
we can write for the LDF wave function
in the rest frame of the particle
\beq
\Phi^{(0)}(x)= \phi^{(0)}({\bf x})e^{-iE_{\bar{q}}t}\ ,\label{vqwf}
\eeq
where $E_{\bar{q}}$ denotes
the energy of the LDF.
The LDF wave function of a meson
moving with (ordinary)
velocity $\bbeta$ along the $z$ axis
(laboratory  frame) is then given by
\beq
\Phi(x') = S(\bbeta) \Phi^{(0)}(x)\ ,\label{labwf}
\eeq
with $x'=\Lambda^{-1}(\bbeta)x$ being the laboratory frame, $x$
the rest frame of the meson, and $S(\bbeta)$ is the wave function
Lorentz boost.

Since IW functions are Lorentz invariant they can
be calculated in any  frame. Particularly
convenient is the Breit frame, in which the two mesons
move with equal and opposite velocities. As noted
in \cite{zalewski2}, the wave functions relevant for the
overlap $\lan \Phi'|\Phi\ran$ are at $t'=0$ in the Breit frame.
Therefore, denoting the three-velocity of the final
meson  as $\bbeta$, by the use of (\ref{labwf})
we have
\bea
\lan \Phi'(v')|\Phi(v)\ran &=& \int d^{3}x' \Phi'^{\dag}(x')
\Phi(x')|_{t'=0}\nonumber \\
&=& \int d^{3}x' \Phi'^{(0)\dag}(x_{+})S^{\dag}(\bbeta)
S(-\bbeta)
\Phi(x_{-})|_{t'=0}\nonumber \\
&=& \int d^{3}x' \Phi'^{(0)\dag}(x_{+})\Phi(x_{-})|_{t'=0}\ .
\label{int}
\eea
In this expression $x_{+}$ and $x_{-}$ denote the rest frames
of the final (moving in the +$z$ direction) and initial meson
(moving in the -$z$ direction), respectively. To obtain
the last equation we have used the  fact
that Lorentz boosts
satisfy $S^{\dag}(\bbeta)= S(\bbeta)=S^{-1}(-\bbeta)$, so that
 boost factors cancel out. Also, we have $(\beta =|\bbeta|$)
\beq
x_{\pm}|_{t'=0} = \Lambda(\pm \bbeta)x'_{t'=0}=
(\mp \gamma \beta z',x',y',\gamma z')\ ,\label{xpm}
\eeq
where, in terms of $\omega$,
\bea
\gamma &= &\sqrt{\frac{\omega+1}{2}}\ ,\label{gbf}\\
\beta &=&\sqrt{\frac{\omega-1}{\omega+1}}\ .\label{bbf}
\eea
Using (\ref{vqwf}) and (\ref{xpm}) in (\ref{int}) we find
\beq
\lan \Phi'(v')|\Phi(v)\ran =
\int d^{3}x' \phi'^{(0)\dag}(x',y',\gamma z')
\phi^{(0)}(x',y',\gamma z')e^{-i(E_{\bar{q}}+E'_{\bar{q}})
\gamma
\beta z'}\ .
\eeq
Finally, after rescaling the $z'$ coordinate ($z'\rar \frac{1}{\gamma}z'$),
renaming integration variables, and using kinematical
identities (\ref{gbf}) and (\ref{bbf}), we obtain
\beq
\lan \Phi'(v')|\Phi(v)\ran =
\sqrt{\frac{2}{\omega+1}}\int d^{3}x
\phi'^{(0)\dag}({\bf x})\phi^{(0)}({\bf x})e^{-iaz}\ ,
\label{ovl}
\eeq
where
\beq
a = (E_{\bar{q}}+E'_{\bar{q}})\sqrt{\frac{\omega-1}{\omega+1}}\ .
\eeq
This formula was first obtained in \cite{zalewski2} for the
semileptonic $B\rar D^{(*)}e\bar{\nu}_{e}$ decays (where
$E'_{\bar{q}}=E_{\bar{q}}$),
and was
already used for the calculation of the corresponding
 Isgur-Wise function
\cite{zalewski2}-\cite{sv3}.

To illustrate the use of the results obtained in
this section, we consider the
non-relativistic quark model as one simple
example.

\section{Non-relativistic quark model}
\label{qmodel}

Assuming that we can describe heavy-light mesons using a simple
non-relativistic potential model, the rest frame LDF wave functions
(with angular momentum $j$ and its projection $\lambda_{j}$),
can be written as
\beq
\phi^{(\alpha L)}_{j\lambda_{j}}({\bf x})= \sum_{m_{L},m_{s}}
R_{\alpha L}(r)Y_{Lm_{L}}(\Omega)\chi_{m_{s}}
\lan L,m_{L};\frac{1}{2},m_{s}|j,\lambda_{j};L,\frac{1}{2}\ran\ ,
\eeq
where $\chi_{m_{s}}$ represent the rest frame spinors normalized
to one, $\chi^{\dag}_{m'_{s}}\chi_{m_{s}}=\delta_{m'_{s},m_{s}}$, and
$\alpha$ represents all other quantum numbers.

Explicitly, taking into account Clebsch-Gordan coefficients, the states
 that we need are
\bea
\phi^{(\alpha 0)}_{\frac{1}{2}\frac{1}{2}}({\bf x}) &=&
R_{\alpha 0}Y_{00}\chi_{\frac{1}{2}}\ ,\label{nr1}\\
\phi^{(\alpha 1)}_{\frac{1}{2}\frac{1}{2}}({\bf x}) &=&
R_{\alpha 1}[\sqrt{\frac{2}{3}}Y_{11}\chi_{-\frac{1}{2}}
-\sqrt{\frac{1}{3}}Y_{10}\chi_{\frac{1}{2}}]\ ,\\
\phi^{(\alpha 1)}_{\frac{3}{2}\frac{1}{2}}({\bf x}) &=&
R_{\alpha 1}[\sqrt{\frac{1}{3}}Y_{11}\chi_{-\frac{1}{2}}
+\sqrt{\frac{2}{3}}Y_{10}\chi_{\frac{1}{2}}]\ ,\\
\phi^{(\alpha 2)}_{\frac{3}{2}\frac{1}{2}}({\bf x}) &=&
R_{\alpha 2}[\sqrt{\frac{3}{5}}Y_{21}\chi_{-\frac{1}{2}}
-\sqrt{\frac{2}{5}}Y_{20}\chi_{\frac{1}{2}}]\ .\label{nr4}
\eea
Now, using the well known expression
\beq
e^{-ikz}=\sum_{l=0}^{\infty}(2l+1)(-i)^{l}j_{l}(kr)
\sqrt{\frac{4\pi}{2 l+1}}Y_{l0}\ ,
\eeq
together with the wave functions given  above, the overlap
expression (\ref{ovl})
gives\footnote{
Using the states analogous to (\ref{nr1})-(\ref{nr4}), but for
$\lambda_{j}=-\frac{1}{2}$, one can  verify that
all overlaps indeed satisfy (\ref{xaxis}) and (\ref{yaxis}), so that
everything  is consistent.}
\bea
\lan \Phi^{\alpha' 0}_{\frac{1}{2}\frac{1}{2}}(v')|
\Phi^{\alpha 0}_{\frac{1}{2}\frac{1}{2}}(v)\ran &=&
\sqrt{\frac{2}{\omega+1}}\lan j_{0}(ar)\ran_{00}^{\alpha'\alpha}\ ,
\label{ovlphi1}\\
\lan \Phi^{\alpha' 1}_{\frac{1}{2}\frac{1}{2}}(v')|
\Phi^{\alpha 0}_{\frac{1}{2}\frac{1}{2}}(v)\ran &=& i
\sqrt{\frac{2}{\omega+1}}
\lan j_{1}(ar)\ran_{10}^{\alpha'\alpha}\ ,\\
\lan \Phi^{\alpha' 1}_{\frac{3}{2}\frac{1}{2}}(v')|
\Phi^{\alpha 0}_{\frac{1}{2}\frac{1}{2}}(v)\ran &=& -i\sqrt{2}
\sqrt{\frac{2}{\omega+1}}
\lan j_{1}(ar)\ran_{10}^{\alpha'\alpha}\ ,\\
\lan \Phi^{\alpha' 2}_{\frac{3}{2}\frac{1}{2}}(v')|
\Phi^{\alpha 0}_{\frac{1}{2}\frac{1}{2}}(v)\ran &=& \sqrt{2}
\sqrt{\frac{2}{\omega+1}}
\lan j_{2}(ar)\ran_{20}^{\alpha'\alpha}\ ,
\label{ovlphi4}
\eea
where
\beq
\lan F(r) \ran_{L'L}^{\alpha'\alpha} = \int r^{2} dr R^{*}_{\alpha'L'}(r)
R_{\alpha L}(r)F(r)\ .
\label{jave}
\eeq

At this point, one can readily obtain the values for the
IW functions (\ref{xic})-(\ref{xig}) and their
derivatives at the zero recoil point. Let us summarize the
results (ignoring irrelevant phase factors and suppressing
quantum numbers $\alpha'$ and $\alpha$):
\begin{itemize}
\item $0_{\frac{1}{2}}^{-}\rar (0_{\frac{1}{2}}^{-},
1_{\frac{1}{2}}^{-})$ transitions.
\bea
\xi_{C}(\omega)& =&
\frac{2}{\omega+1}
\lan j_{0}(ar)\ran_{00} \ ,\label{xicq}\\
\xi_{C}(1)& =& \lan 1\ran_{00}\ ,\\
\xi'_{C}(1) &=& -\frac{1}{2} - \frac{1}{12}
(E_{\bar{q}}+ E'_{\bar{q}})^{2}<r^{2}>_{00}\ .\label{xicsl}
\eea
Note that these expressions include transitions from the ground state
into radially
 excited states. If the two $j=\frac{1}{2}$ states are the same,
$\xi_{C}(1)$ is normalized to one and $E'_{\bar{q}}=E_{\bar{q}}$.
\item$0_{\frac{1}{2}}^{-}\rar (0_{\frac{1}{2}}^{+},
1_{\frac{1}{2}}^{+})$ transitions.
\bea
\xi_{E}(\omega)& =&
\frac{2}{\sqrt{\omega^{2}-1}}
\lan j_{1}(ar)\ran_{10}\label{xieq}
 \ ,\\
\xi_{E}(1)& =& \frac{1}{3}(E_{\bar{q}}+E'_{\bar{q}})\lan r\ran_{10}\ ,\\
\xi'_{E}(1) &=& -\frac{1}{6}(E_{\bar{q}}+E'_{\bar{q}})\lan r\ran_{10}
 - \frac{1}{60} (E_{\bar{q}}+E'_{\bar{q}})^{3}<r^{3}>_{10}\ .
\eea
\item$0_{\frac{1}{2}}^{-}\rar (1_{\frac{3}{2}}^{+},
2_{\frac{3}{2}}^{+})$ transitions.
\bea
\xi_{F}(\omega)& =&
\sqrt{\frac{3}{\omega^{2}-1}}\frac{2}{\omega+1}
\lan j_{1}(ar)\ran_{10} \label{xifq}\ ,\\
\xi_{F}(1)& =& \frac{1}{2\sqrt{3}}
(E_{\bar{q}}+E'_{\bar{q}})\lan r\ran_{10}\ ,\\
\xi'_{F}(1) &=& -\frac{1}{2\sqrt{3}}(E_{\bar{q}}+E'_{\bar{q}})\lan r\ran_{10}
 - \frac{1}{40\sqrt{3}} (E_{\bar{q}}+E'_{\bar{q}})^{3}<r^{3}>_{10}\ .
\label{xigq}
\eea
\item$0_{\frac{1}{2}}^{-}\rar (1_{\frac{3}{2}}^{-},
2_{\frac{3}{2}}^{-})$ transitions.
\bea
\xi_{G}(\omega)& =&
\frac{2\sqrt{3}}{\omega^{2}-1}
\lan j_{2}(ar)\ran_{20} \ ,\\
\xi_{G}(1)& =& \frac{1}{10\sqrt{3}}
(E_{\bar{q}}+E'_{\bar{q}})^{2}\lan r^{2}\ran_{20}\ ,\\
\xi'_{G}(1) &=& -\frac{1}{10\sqrt{3}}(E_{\bar{q}}+E'_{\bar{q}})^{2}
\lan r^{2}\ran_{20}
 - \frac{1}{280\sqrt{3}} (E_{\bar{q}}+E'_{\bar{q}})^{4}<r^{4}>_{20}\ .
\label{xigdq}
\eea
\end{itemize}

Let us briefly compare these results with the formulation
of \cite{mannel}, which was also followed by \cite{ahmady}. There, apart
from  irrelevant phase factors, different
IW functions are identified directly as overlaps of the wave functions
of the initial and the final state of the LDF
in the rest frame of the initial meson. Explicitly (putting a
tilde over the
form factors to avoid confusion),
\bea
\tilde{\xi}_{C}(\omega)& =& \lan j_{0}(\tilde{a}r)\ran_{00}\ ,
\label{xiec}\\
\tilde{\xi}_{E}(\omega)& =& \sqrt{3}\lan j_{1}(\tilde{a}r)\ran_{10}\ ,
\label{xiem}\\
\tilde{\xi}_{F}(\omega)& =& \sqrt{3}\lan j_{1}(\tilde{a}r)\ran_{10}\ ,
\label{xifm}\\
\tilde{\xi}_{G}(\omega)& =& \sqrt{5}\lan j_{2}(\tilde{a}r)\ran_{20}\ ,
\label{xieg}
\eea
with the definition
\beq
\tilde{a}=E'_{\bar{q}}\sqrt{\omega^{2}-1}\ .
\eeq
This formulation has several difficulties.
For example, if  one uses harmonic oscillator wave functions
(as was done in \cite{mannel})
in order to estimate overlaps $\lan j_{j}(\tilde{a}r)\ran $, then it
is easy to show that
\beq
\tilde{\xi}'_{C}(1)=-\frac{E'^{2}_{\bar{q}}}{2\beta^{2}_{B}}\ ,
\eeq
where $\beta_{B}$ is the variational parameter. Using the values
from the ISGW model \cite{isgw} (which is also used in \cite{mannel}),
$\beta_{B}\approx 0.4\ GeV$ and $E'_{\bar{q}}\approx 330\ MeV$,
one gets the slope of the IW function for the semileptonic $B$
decays,
\beq
\tilde{\xi}'_{C}(1) \approx -0.34\ ,
\eeq
which is clearly too large \cite{cleo}. On the other hand,
the same wave function used in (\ref{xicsl}) gives ($E_{\bar{q}}=
E'_{\bar{q}}\approx 330\ MeV$)
\beq
\xi'_{C}(1) \approx -0.84\ ,
\eeq
a result which is in much better agreement with the data \cite{cleo}.
Furthermore, from (\ref{xiec})-(\ref{xieg}) one can see that
all form factors (except for $\tilde{\xi}_{C}$) vanish at the
zero recoil point,
whereas the Bjorken sum rule \cite{bjorken} requires a nonvanishing
$P$-wave form
factor in this limit. It is clear that our form factors
(\ref{xieq}) and (\ref{xifq}) do not suffer from that problem.
Also, from (\ref{xiem}) and (\ref{xifm}) one can see that
$\tilde{\xi}_{E}(\omega)=\tilde{\xi}_{F}(\omega)$,
while (\ref{xieq}) and (\ref{xifq}) imply that our form factors satisfy
\beq
\xi_{E}(\omega) = \frac{\omega+1}{\sqrt{3}}\xi_{F}(\omega)\ ,
\eeq
and in particular
\beq
\xi_{E}(1)= \frac{2}{\sqrt{3}}\xi_{F}(1)\ .
\eeq
Even though we shall postpone a full account of the radiative
rare $B$ decays for later \cite{rareb}, let
 us just point out that the authors of \cite{mannel}
emphasize a substantially larger branching fraction for the decay
$B\rar K_{2}^{*}(1430)\gamma$ than the one found in
\cite{altomari}. If
one includes only the factor of
$\sqrt{\frac{3}{\omega-1}}\frac{1}{\omega+1}$ from (\ref{fxi})
(which is about $\frac{1}{\sqrt{3}}$ for $\omega\approx 2$), one
gets a factor of $3$ smaller result than the one quoted in \cite{mannel},
bringing this particular branching ratio into much better
agreement with the result of \cite{altomari}.

Finally, we  can generalize the
quark model  approach to any model involving the
Dirac equation with a spherically symmetric potential.
There, the wave function has the form
\beq
\phi_{j\lambda_{j}}^{(\alpha k)}({\bf x}) =
\left( \begin{array}{c}
f_{\alpha j}^{ k}(r) {\cal Y}_{j\lambda_{j}}^{k}(\Omega)\\
i g_{\alpha j}^{ k}(r) {\cal Y}_{j\lambda_{j}}^{-k}(\Omega)\end{array}
\right)\ ,
\eeq
where ${\cal Y}_{j\lambda_{j}}^{k}$ are the usual
spherical spinors,
 $k=l$ ($l=j+\frac{1}{2}$) or $k=-l-1$ $(l=j-\frac{1}{2})$, and
$\alpha$ again denotes all other quantum numbers. Using this it can be
shown that all the expressions (\ref{ovlphi1})-(\ref{ovlphi4}) and
(\ref{xicq})-(\ref{xigdq}) are unchanged, except that the
expectation value (\ref{jave}) is
replaced by
\beq
\lan F(r) \ran_{L'L}^{\alpha'\alpha}\rar
\lan F(r) \ran_{j'j}^{\alpha'\alpha} = \int r^{2} dr
[f^{*k'}_{\alpha' j'}(r)f^{k}_{\alpha j}(r) +
g^{*k'}_{\alpha' j'}(r)g^{k}_{\alpha j}(r) ] F(r)\ .
\eeq

\section{Conclusions}
\label{conc}

We have presented a simple method for relating
form factors as defined by the covariant
trace formalism \cite{georgi2,korner,falk}, to the explicit
overlaps of the rest frame wave functions describing the initial and
the final states of the light degrees of freedom. We have
obtained explicit formulae for several transitions of interest
(from $0^{-}$ into a few lowest excited states), and have shown
how one can apply these expressions in the simple
quark model, and in models involving the Dirac equation with spherically
symmetric potentials.
We have also pointed out several inconsistencies present in
recent calculations of radiative rare $B$ decays into higher
$K$-resonances \cite{mannel,ahmady},
and have shown how these can be fixed even within the same non-relativistic
quark model that was used in \cite{mannel}. A full account
of the radiative rare $B$ decays will be presented elsewhere
\cite{rareb}.

\vskip 1cm
\begin{center}
ACKNOWLEDGMENTS
\end{center}
We would like to thank J. F. Amundson for several useful
conversations.
This work was supported in part by the U.S. Department of Energy
under Contract No. DE-FG02-95ER40896 and in part by the University
of Wisconsin Research Committee with funds granted by the Wisconsin Alumni
Research Foundation.

\end{document}